\documentclass[referee,pdflatex,sn-nature,super]{sn-jnl}

\usepackage{graphicx}%
\usepackage{multirow}%
\usepackage{fix-cm}

\usepackage{amsmath,amssymb,amsfonts}%
\usepackage{amsthm}%
\usepackage[title]{appendix}%
\usepackage{xcolor}%
\usepackage{textcomp}%
\usepackage{manyfoot}%
\usepackage{booktabs}%
\usepackage{algorithm}%
\usepackage{algorithmicx}%
\usepackage{algpseudocode}%
\usepackage{listings}%
\usepackage[capitalise]{cleveref}
\usepackage[version=4]{mhchem}
\usepackage{braket}
\usepackage[final]{changes}            
\usepackage{xcolor}
\usepackage{xpatch}
\usepackage{todonotes}
\makeatletter
\def\changeBibColor#1{
  \in@{#1}{}
  \ifin@\color{red}\else\normalcolor\fi
}
 
\xpatchcmd\@lbibitem
  {\item}
  {\changeBibColor{#2}\item}
  {}{\fail}
\makeatother
\unnumbered

\begin{document}

\title[Enhancement of Quantum Coherence]{Enhancement of quantum coherence in solid-state qubits via interface engineering }

\author[1]{\fnm{Wing Ki} \sur{Lo}}
\equalcont{These authors contributed equally to this work.}
\author[1]{\fnm{Yaowen} \sur{Zhang}}
\equalcont{These authors contributed equally to this work.}
\author[1]{\fnm{Ho Yin} \sur{Chow}}
\equalcont{These authors contributed equally to this work.}
\author[1]{\fnm{Jiahao} \sur{Wu}}
\equalcont{These authors contributed equally to this work.}
\author[1]{\fnm{Man Yin} \sur{Leung}}
\author[1]{\fnm{Kin On} \sur{Ho}}
\author[1]{\fnm{Xuliang} \sur{Du}}
\author[1]{\fnm{Yifan} \sur{Chen}}
\author[1]{\fnm{Yang} \sur{Shen}}

\author*[1,2]{\fnm{Ding} \sur{Pan}}\email{dingpan@ust.hk}

\author*[1]{\fnm{Sen} \sur{Yang}}\email{phsyang@ust.hk}

\affil[1]{\orgdiv{Department of Physics}, \orgname{The Hong Kong University of Science and Technology}, \orgaddress{\street{Clear Water Bay, Kowloon}, \city{Hong Kong}, \country{China}}}

\affil[2]{\orgdiv{Department of Chemistry}, \orgname{The Hong Kong University of Science and Technology}, \orgaddress{\street{Clear Water Bay}, \city{Hong Kong}, \country{China}}}

\abstract{
Shallow nitrogen-vacancy (NV) centers in diamond are promising quantum sensors but suffer from noise-induced short coherence times due to bulk and surface impurities. We present interfacial engineering via oxygen termination and graphene patching, extending shallow NV coherence to over 1 ms, approaching the \ce{T_{1}} limit. Raman spectroscopy and density-functional theory reveal surface termination-driven graphene charge transfer reduces spin noise by pairing surface electrons, supported by double electron-electron resonance spectroscopy showing fewer unpaired spins. Enhanced sensitivity enables detection of single weakly coupled \ce{^{13}C} nuclear spins and external \ce{^{11}B} spins from a hexagonal boron nitride (h-BN) layer, achieving nanoscale nuclear magnetic resonance. A protective h-BN top layer stabilizes the platform, ensuring robustness against harsh treatments and compatibility with target materials. This integrated approach advances practical quantum sensing by combining extended coherence, improved sensitivity, and device durability.
}

\keywords{Graphene, Solid-state qubits, Interface engineering, Coherence time}

\maketitle

\section{Introduction}\label{Intro}
Solid-state qubits are one of the most promising quantum sensors due to their sensitivity and resolution at the nanoscale \cite{PhysRevApplied.19.044086, Webb2019, Cochrane2016}. By fabricating shallow qubits near the surfaces, one can realize a nanoscale sample-to-sensor distance \cite{Watanabe2021, Neethirajan2023, Maze2008}, which greatly enhances the coupling strength between them and eventually reaches the sensitivity of a single proton limit \cite{doi:10.1126/science.aad8022}. However, these shallow solid-state qubits commonly exhibit reduced coherence time due to the influence of surrounding spin baths, limiting the sensitivity of quantum sensing \cite{CHROSTOSKI2021412767}. The spin baths associated with shallow solid-state qubits originate from various nuclear and electron spins within the bulk and on the surface. Taking nitrogen-vacancy (NV) centers in diamonds as an example, various strategies have been employed to mitigate the spin noise within the bulk \cite{Itoh2014, FávarodeOliveira2017}.

Nonetheless, it is very difficult to reduce the surface electron spins, which are unavoidable {due to the unpaired electrons on the diamond surface}. Unpaired surface electron spins induce noise sources such as proximal spin noise and electric noise \cite{CHROSTOSKI2021412767, PhysRevLett.118.197201, D2TC01258H, PhysRevLett.112.147602, PhysRevB.86.081406, PhysRevLett.114.017601} and dominate the decoherence mechanism of the NV center especially for shallow NV centers (1 - 20 nm). Hence, it remains challenging to improve the quality of shallow qubits. Up to now, the stability and coherence time of shallow NV centers have been very limited \cite{D2TC01258H, PhysRevLett.112.147602, PhysRevLett.114.017601, PhysRevLett.118.197201}. To address this issue, researchers have explored different approaches, including modulation of the electronic spin bath by atomic force microscope (AFM) through electric field manipulation \cite{Zheng2022} and covering the diamond surface with liquids of high dielectric constant \cite{PhysRevLett.115.087602}.
However, special instrumentation and strong local electric field created by these methods hinder their compatibility with general quantum sensing applications.

Drawing inspiration from band alignment modifications achieved through tailored surface and interface functionalization of semiconducting materials in photovoltaic applications\cite{doi:10.1021/jp3124583, doi:10.1021/ja5079865}, 
we propose an interface engineering method using graphene patching on the oxygen-terminated (O-terminated) diamond to reduce the spin noise from surface unpaired electrons \cite{Zheng2022,10.1093/nsr/nwaf076} and enhance the coherent time of shallow NV centers, which can be readily adapted to additional sensing applications. Systematic characterization of 20 shallow NV centers revealed coherence enhancement across all NVs. The alternating current (AC) magnetic field sensitivity of shallow NV centers is enhanced to $23 \ \mathrm{nT/Hz^{1/2}}$, approaching the best sensitivity (9 $\mathrm{nT/Hz^{1/2}}$) reported for deep NV centers in an enriched \ce{^{12}C} and \ce{^{28}Si} diamond with phosphorus-doping \cite{Herbschleb2019}. We use Carr–Purcell–Meiboom–Gill (CPMG) sequence to further extend the coherence time of shallow NV centers to greater than 1 ms, approaching the $\mathrm{T_1}$ limit ($1.6 \pm  0.3 \ \mathrm{ms}$) of NV centers.(see Supplementary Note 1) 

To demonstrate quantum sensing capabilities of our coherence-enhanced shallow NV centers, we perform the sensing of both weak \ce{^{13}C} nuclear spins inside diamond and \ce{^{11}B} nuclear spin baths of the hexagonal boron nitride (h-BN) outside the diamond. Unlike internal spins, probing external spins imposes extreme challenges due to their weaker coupling and rapid decoherence. Achieving such sensitivity requires not only exceptional spin coherence but also an extreme close distance between the NV center and the target spins, a feat previously considered near-impossible without purified diamond sample and vast selected NV center. 
Using our interface engineering approach, we demonstrate external spin sensing with shallow NV centers in non-purified diamond, achieving nanoscale NMR of molecular structures, quantum materials, and biological systems at the atomic scale. Our findings highlight the potential of this interface engineering method for improving the quantum performance of shallow solid-state qubits. 


The surface engineering method utilizes the gapless energy band structure of graphene as an electron source to pair with the unpaired electrons on the O-terminated diamond surface. To elucidate the physical mechanisms underlying this interface engineering method, we employed double electron-electron resonance (DEER) spectroscopy, density functional theory (DFT) calculations, and Raman spectroscopy. DEER spectroscopy of shallow NV centers exhibits at least an order of reduction in the unpaired electron spin concentration, approximately $10^{11}$ cm$^{-2}$, after interface engineering. Raman spectra of graphene indicate hole doping with a density on the order of $\sim10^{12}$ cm$^{-2}$ when transferred to an O-terminated diamond surface. Independent DFT calculations support this finding, showing that the O-termination has a lower Fermi level and facilitates electron transfer from graphene to {pair with the unpaired electrons of carbon atoms on the diamond surface}. It should be noted that we reach a quantitative agreement without any fitting parameters. 

\begin{figure}[t]
\includegraphics[width=\textwidth]{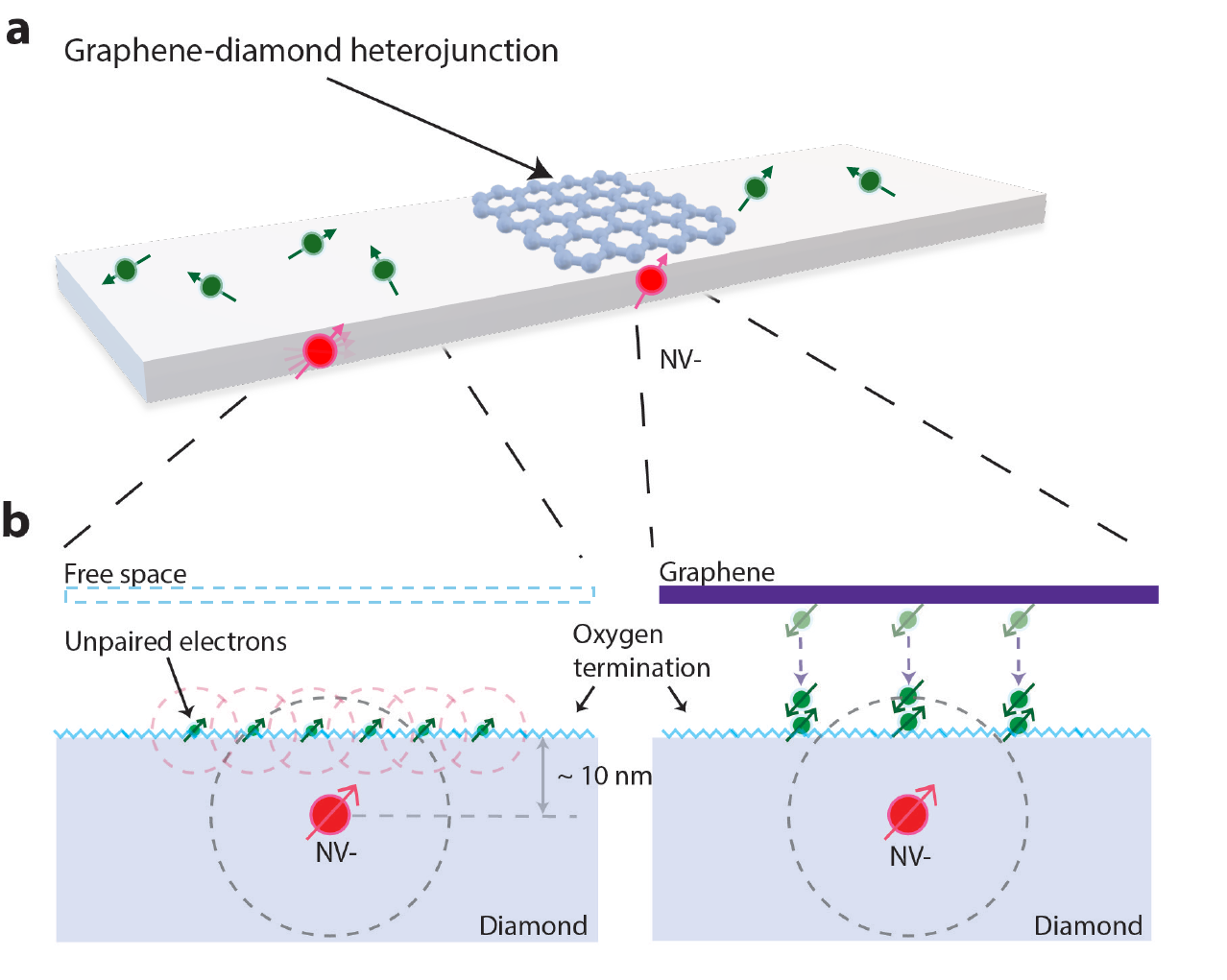}
\caption{\textbf{An illustration of the interface engineering method and the proposed charge transfer mechanism.} \textbf{a}, Schematic showing an NV center near the diamond surface. The surface has {unpaired electrons} (green arrows) that produce spin noise. At the graphene O-terminated diamond heterojunction, there is less noise fluctuation on the NV center. \textbf{b}, An illustration showing how unpaired electrons on the diamond surface cause decoherence on the shallow NV center under an external magnetic field. Upon patching graphene on the O-terminated surface, electrons transfer from graphene towards the diamond surface, pairing up the surface electrons.}
\label{fig1}
\end{figure}

\section{Results}
\subsection{Quantum performance enhancement of shallow NV centers}\label{T2}

\begin{figure}[t]
\includegraphics[width=\textwidth]{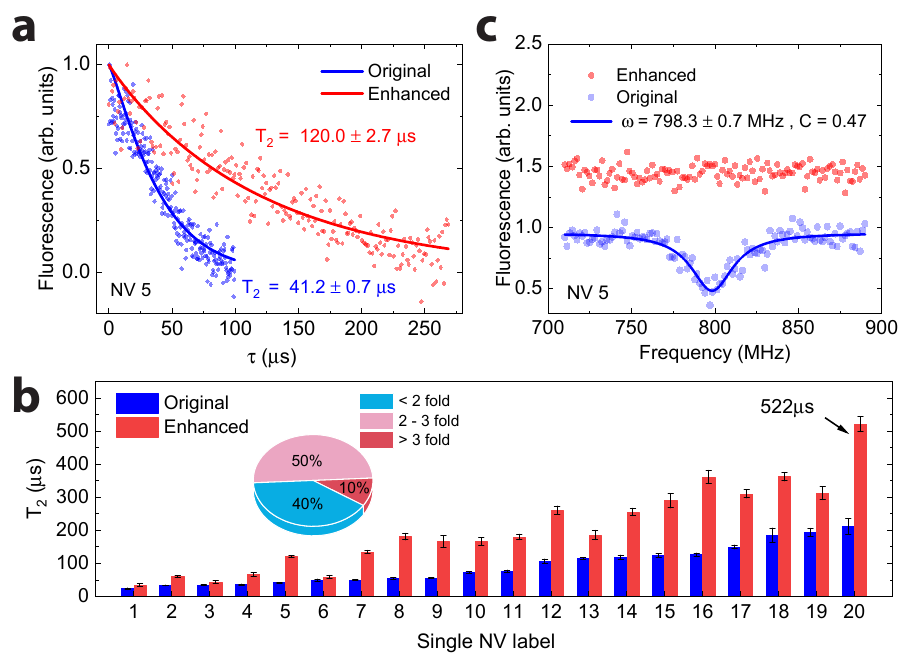}
\caption{\textbf{Comparison of the coherence time and unpaired electron spins between an O-terminated diamond and graphene-diamond heterojunction.} \textbf{a}, Hahn echo decay curves from the same NV center. The Hahn echo coherence time $\mathrm{T_{2}}$ is improved $\sim$3 fold after graphene patching. \textbf{b}, Summary of the Hahn echo coherence time $\mathrm{T_{2}}$ enhancement of all 20 shallow NVs after graphene patching. They are arranged in ascending order of the $\mathrm{T_{2}}$ before enhancement. Error bars represent the fitting uncertainty. \textbf{c}, DEER spectroscopy detected by NV5. Blue dots denote the surface unpaired electron signals resonated at $ \omega = \gamma_e B = 798 $ MHz, where C is the contrast. The surface unpaired electron spins signal disappear after graphene patching, denoted by red dots. The red and blue curves are offset for clarity. }
\label{fig2}
\end{figure}

To create an O-terminated diamond, the diamond sample is cleaned with triacid boiling \cite{BROWN201940, PhysRevX.9.031052}(see Supplementary Note 2). Triacid treatment is commonly applied for the NV center experiment as surface oxidation helps stabilize the NV negative charged state. However, the unpaired electron due to the imperfectness of the diamond surface remains a critical spin noise source. To resolve those surface spin noise, we transfer graphene to cover the entire implanted diamond surface following the transfer procedure (see methods) immediately after the triacid boiling. The resultant graphene-diamond heterojunction is illustrated in \cref{fig1}a. The graphene provides electrons to pair the surface unpaired electrons to remove spin noise as shown in \cref{fig1}b.

We perform single-spin measurements on 20 shallow shallow NV centers approximately 5 - 20 nm (see Supplementary Note 1) in the depth of an implanted (100)-O-terminated diamond and compare their performance with and without graphene. Hahn-echo ($\mathrm{T_{2}}$) measurements of NV centers are highly sensitive to the surrounding noise spectrum, thus it is ideal to probe the changes in nearby spin baths. \cref{fig2}a shows a comparison of the Hahn echo decay of the same NV center ($\sim$14 nm below the diamond surface) before and after the graphene patching. The measurements are performed at 286 G. We fit the Hahn-echo signal to a stretched exponential $e^{-(\tau/T_{2})^{p}}$ that takes into account the spin bath correlations \cite{PhysRevB.87.115122}, where $\mathrm{p}$ is a fitting parameter. The $\mathrm{T_{2}}$ time has a significant enhancement upon graphene transfer. The improvement is observed in all NV centers ranging from 1.2 to at most 3.3 fold, as summarized in \cref{fig2}b. Notably, the coherence time for NV20 extended to as long as 522 $\mu$s, indicating the noise is strongly suppressed with graphene patching. Compared with state-of-the-art techniques for enhancing $\mathrm{T_2}$ in shallow NV centers , including electric field manipulation (maximum $\sim$ 170 $\mathrm{\mu s}$)\cite{Zheng2022} and surface termination (maximum $\sim$ 70 $\mathrm{\mu s}$)\cite{PhysRevX.9.031052}, our approach achieves a longer coherence times (maximum 522 $\mathrm{\mu s}$) approaching the bulk diamond limit while maintaining both simplicity and stable performance. This represents a significant improvement for shallow NV applications in quantum sensing.
For $\mathrm{T_2}^*$, we observed no significant enhancement due to low-frequency spin noise induced by laser illumination\cite{PhysRevLett.122.076101, Lozovoi2021}, which lies outside the scope of our noise elimination targets. 

To confirm unpaired electron spins as the dominant spin noise on the diamond surface, we performed DEER measurements on 7 shallow NV centers (NV 1,2,3,5,7,16\&17) and probed the proximal spin baths. By driving the corresponding resonance frequency, this method is able to measure surface spins \cite{PRXQuantum.3.040328,10.1093/nsr/nwaf076,Grotz_2011,PhysRevB.86.195422,Zheng2022} and other qubits in the bulk diamond lattice \cite{PhysRevB.104.094307, Degen2021}. The unpaired electrons from the diamond surface carbon atoms will precess along our applied magnetic field and create spin noise to the NV center. Under an applied 286 G magnetic field, the precession of unpaired electron spins results in a radio frequency (RF) at $ \omega = \gamma_e B = 798 $ MHz, where $ \gamma_e= 2.8 $ MHz/G is the gyromagnetic ratio of a free electron. Since 98.9 \% of carbon atoms are \ce{^{12}C} which is spinless, the unpaired electrons from carbon atoms will precess at Larmor frequency without hyperfine interactions. The DEER signal $S(\omega)$ is fitted with the Lorentzian function as

\begin{align}
S(\omega) = 1 - C\frac{\Delta\omega^{2}}{(\omega-\omega_{0})^{2}-\Delta\omega^{2}},
\end{align}
where $\mathrm{C}$ is the contrast, $\mathrm{\omega_{0}}$ is the center of the Lorentzian, and $\Delta \omega$ is the half width at half maximum (HWHM).

The DEER spectra are shown in \cref{fig2}c. For the spectrum of an O-terminated diamond, the fitted resonance frequency is $\omega_{0} = 798.3 \pm 0.7 $ MHz and the HWHM $\Delta\omega=15.4  \pm 1.4$ MHz. This shows noticeable surface spins with an O-terminated diamond surface. Upon the graphene transfer, in sharp contrast, the unpaired electron spins signal is no longer detectable. To further investigate this effect, we conduct DEER decay measurements which have previously demonstrated to estimate the concentrations of unpaired electron spins on the diamond surface \cite{PRXQuantum.3.040328}. For the 7 single NVs, an average of 0.72 $\times10^{11}$ cm$^{-2}$ are observed (see Supplementary Note 1). The resonance frequency of the DEER spectrum confirms that the dominant sources of the spin noise come from the unpaired electron spin baths. Therefore, the enhancement in the $\mathrm{T_{2}}$ time and the reduction of the unpaired electron spins DEER signal confirm the pairing of unpaired electrons with the use of graphene, revealing the significant effect of proposed interface engineering.

\subsection{Studying the charge transfer mechanism with Raman spectra}\label{Raman}

\begin{figure}[t]
\includegraphics[width=\textwidth]{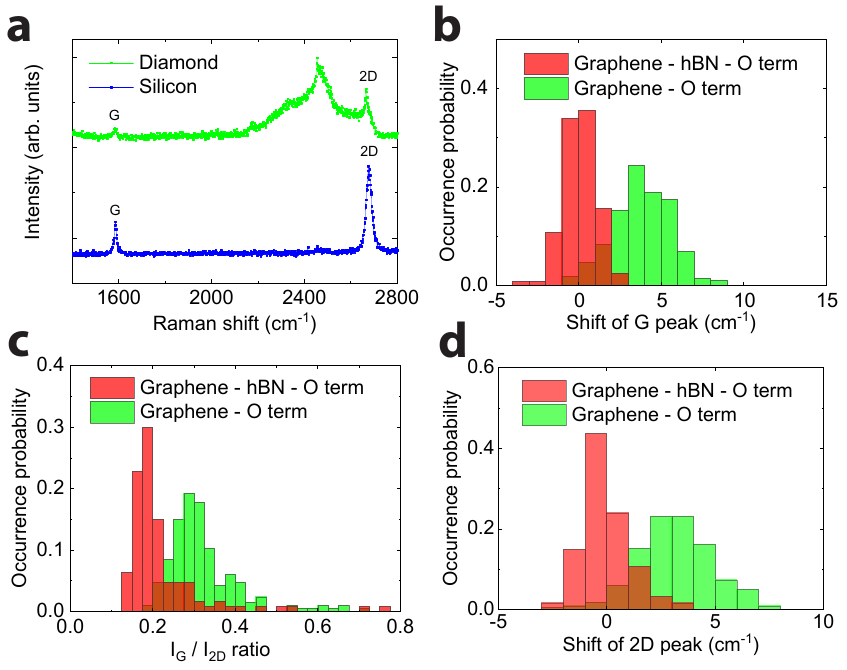}
\caption{\textbf{Shift and intensity of Raman spectra in graphene indicate charge transfer in graphene O-terminated diamond heterojunction.} \textbf{a}, Raman spectrum showing the characteristic G and 2D bands of graphene on silicon and diamond substrate after a standard transfer process. \textbf{b}, A statistic trend of the relative G band shift of graphene in direct contact with the O-terminated diamond (green) and with h-BN buffer (orange). The graphene on the h-BN flake at 1584 cm$^{-1}$ is set as the 0 shift reference. \textbf{c}, A statistic trend of the relative 2D band shift of graphene in direct contact with the O-terminated diamond (green) and with h-BN buffer (orange). The graphene on the h-BN flake at 2683 cm$^{-1}$ is set as the 0 shift reference. \textbf{d}, A statistical trend of the relative intensity $\mathrm{I_G/I_{2D}}$ ratio in direct contact with O-terminated diamond (green) and with h-BN buffer (orange). Histograms are collected by Raman mapping analysis over a graphene transfer region with around 200 data.}
\label{fig3}
\end{figure}

To confirm the occurrence of charge transfer on the graphene, we use Raman spectroscopy, a nondestructive optical probe, to study the changes in the electronic properties of graphene. In \cref{fig3}a, two characteristic G and 2D bands are measured after a standard transfer procedure (see methods for details) on silicon and diamond samples, respectively. The positions of the bands are centered at $\sim$ 1585 cm$^{-1}$ for G and $\sim$ 2685 cm$^{-1}$ for 2D of the silicon sample in excellent agreement with previous studies \cite{doi:10.1126/science.1171245} that indicate a successful graphene transfer. Since the graphene 2D band overlaps with the second-order Raman band of a diamond, the 2D band positions and intensities in our subsequent analysis are obtained by subtracting the signal from the same diamond without graphene while keeping the same measurement parameters. 

To study the influences of O-terminated diamond acting on graphene, we introduce some uniform hexagonal boron nitride (h-BN) flakes (see Supplementary Note 3) between the graphene and diamond interfaces. The h-BN serves as an atomically flat and decreased native charge density platform to probe the intrinsic Raman signal of graphene \cite{Dean2010}. To avoid sample inhomogeneity, we carry out Raman mapping analysis to obtain the band shift statistically. The characteristic band (1584 cm$^{-1}$ for G, 2683 cm$^{-1}$ for 2D) of graphene on a uniform h-BN sample is set as a reference point \cite{wang2012negligible} for comparison with diamond. In \cref{fig3}b, graphene directly in contact with the O-terminated sample shows $\sim$ 3.8 cm$^{-1}$ blue shift of the G band relative to the h-BN one. This shift is unlikely to be attributed to surface impurities and roughness, as these factors typically alter the phonon velocity of graphene, resulting in a red shift in the G band\cite{spear2015influence}. Previous studies showed that the charge doping of graphene can induce G band blue shift \cite{kalbac2010influence,stampfer2007raman,Pisana2007,PhysRevLett.98.166802,FERRARI200747}. The increase of the intensity ratio $\mathrm{I_G/I_{2D}}$ shown in \cref{fig3}c also suggests an occurrence of charge doping in graphene\cite{Das2008} when it is in direct contact with O-terminated diamond. The charge doping changes the Fermi surface, which moves the Kohn anomaly away from the phonon wavevector $\mathbf{q}=0$, leading to a stiffening of the G band \cite{Pisana2007}. The h-BN data remain a good reference due to its high dielectric constant\cite{Watanabe2004} that the flakes serve as charge-insulating layers between the diamond surface and graphene, prohibiting charge transfer. Moreover, the blue shift in the 2D band showed in \cref{fig3}d indicates hole doping \cite{Pisana2007, PhysRevLett.97.266407, Das2008}, suggesting net electron export in the graphene layer. Therefore, we deduce that hole doping occurs at graphene directly patched to the O-terminated diamond, which is not observed when an insulating h-BN flake is interposed. 

The doping-induced Fermi level shift from the Dirac point can be estimated using the relation:
\begin{align}
    \epsilon_F = \text{sgn}(n)\sqrt{|n|\pi}\hbar v_F,
    \label{eq:vF_n}
\end{align}
where $n$ denotes the doping concentration, $\text{sgn}(\cdot)$ returns the sign of the variable and $v_F$ represents the Fermi velocity ($\hbar v_F$ = 5.52 eV\AA \cite{Pisana2007}). Pisana \textit{et al.} \cite{Pisana2007} and Lazzeri \textit{et al.} \cite{PhysRevLett.97.266407} further give the doping-induced G band shift relation:
\begin{align}
      \hbar \Delta \omega = \alpha' |\epsilon_F| + \frac{\alpha' \hbar \omega_{0}}{4}\ln\left(\frac{|\epsilon_F| - \hbar \omega_0/2}{|\epsilon_F| + \hbar \omega_0/2}\right),
      \label{eq:G_shift}
\end{align}
where $\omega_0$ is the frequency of G band and $\alpha'$ is determined by 
\begin{align}
   \alpha' = \frac{\hbar A \langle D^2_\Gamma \rangle_F}{\pi M \omega_0 (\hbar v_F)^2}.
   \label{eq:alpha}
\end{align}
Here $\mathrm{A}$ denotes unit cell area, $ \langle D^2_\Gamma \rangle_F$ denotes the deformation potential of the $E_{2g}$ mode, $\hbar$ is the reduced Planck constant, and $\mathrm{M}$ is the carbon mass. When there is a $\sim3.8$ cm$^{-1}$ G band shift, the hole doping concentration is estimated using Eqs. (\ref{eq:vF_n}, \ref{eq:G_shift}, and \ref{eq:alpha}) as $\sim  10^{12} $ cm$^{-2}$, which is larger than the result observed in the DEER measurement ($0.72$ $\times 10^{11}$ cm$^{-2}$).
The shift in graphene’s Raman peak signifies electron transfer from graphene to the diamond surface. The majority of these electrons pair with dangling bonds to suppress surface spin noise, while the remaining fraction could be captured by surface adsorbates, such as water molecules\cite{doi:10.1021/jp910971e,doi:10.1126/science.1148841}. DEER measurements selectively probe the unpaired electron spin bath linked to \ce{^{12}C} and oxygen atoms that cannot capture the surface absorption events. The charge transfer values derived from Raman measurements correlate well with the defect concentration of the O-terminated surface obtained by  density functional theory and synchrotron-based X-ray absorption spectroscopy\cite{https://doi.org/10.1002/admi.201801449}.

\subsection{DFT calculation}\label{DFT}
To investigate the atomistic mechanisms of interfacial engineering, 
we perform DFT calculations on the heterojunctions of graphene on a (100)-O-terminated diamond surface [O\_D(100)], and also (100)-hydroxyl-terminated (OH-terminated) diamond surface [OH\_D(100)]. These two surface terminations are 
representative in our NV measurements (see the Method section), 
and previous experimental studies \cite{PhysRevX.9.031052, LI2021725}. The diamond surfaces are functionalized by the functional groups C-O-H, C=O and C-O-C. The concentration of C-O-H, C=O, and C-O-C in OH\_D(100) are 50\%, 25\%, and 25\%, respectively. The concentration of C-O-H, C=O, and C-O-C in O\_D(100) are 16.7\%, 33.3\%, and 50\%, respectively. Here, these concentrations are calculated as the ratio of the number of each specific functional group to the total number of functional groups on the surfaces.
There is one unpaired electron in one carbon atom in the O\_D(100) surface. 
Such unpaired electrons
are the primary source of spin noise. Considering our simulation lattice of $2.46  \text{ \AA} \times 17.05 \text{ \AA} $, the concentration of unpaired electrons is  $2.4 \times 10^{14} \text{ cm}^{-2}$, which is four orders of magnitude larger than the concentration in our DEER measurement and also the value estimated by Eq. (\ref{eq:vF_n}-\ref{eq:alpha}). 
 In the DEER measurement, the unpaired electron concentration is $0.72$ $\times 10^{11} \text{cm}^{-2}$.
Reproducing a low unpaired electron concentration as found in the DEER measurement requires a much larger computational cell ($1.6 \times 10^5 \text{\AA}^2$), which exceeds our current computational capabilities. Here, we used a small 
unit cell to study the mechanism of charge transfer.

\begin{figure}[htbp]
    \centering
    \includegraphics[width=\linewidth]{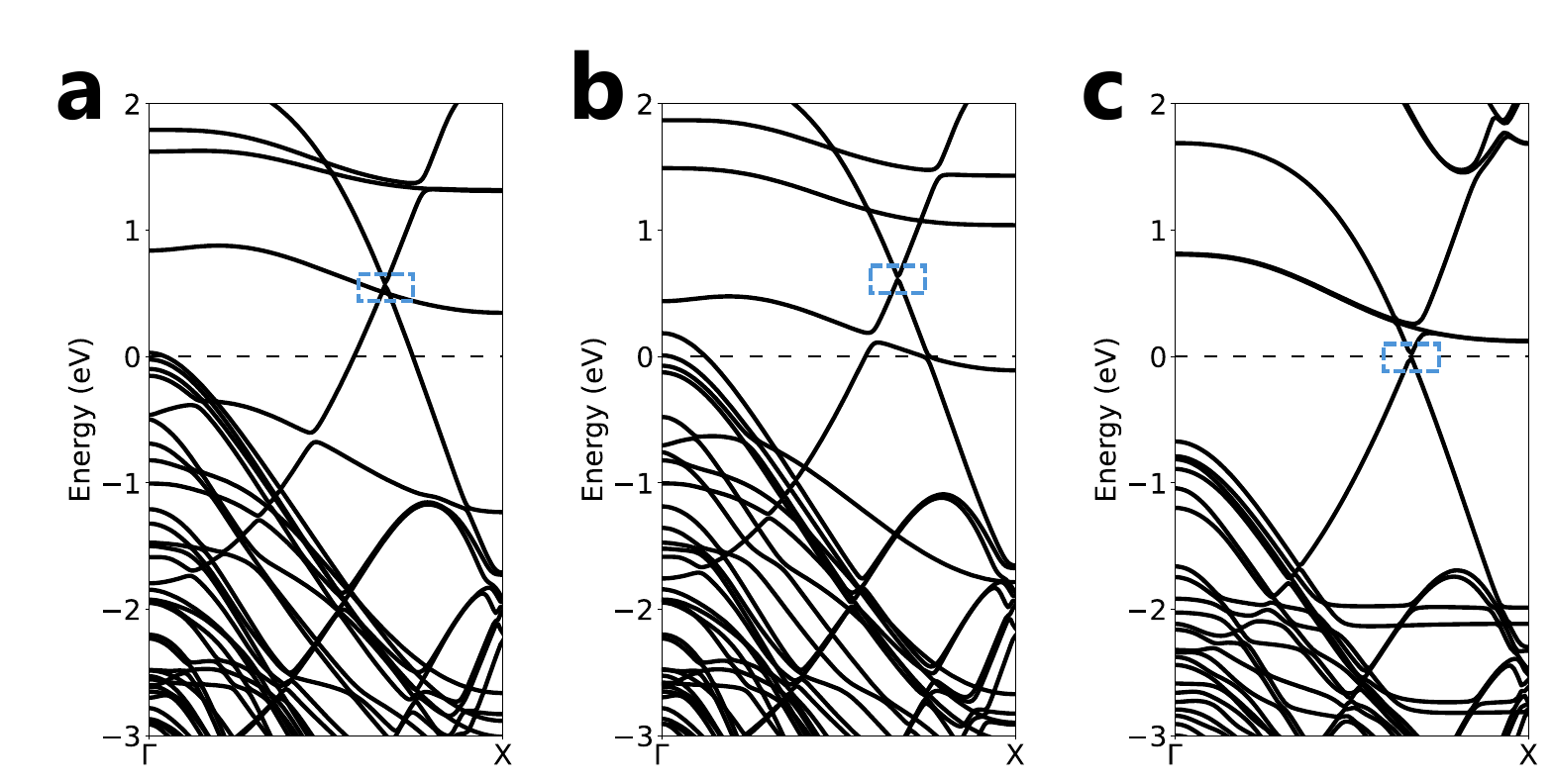}
     \caption{\textbf{DFT calculation on graphene-diamond heterojunctions.} Electronic band structure of the heterojunction G/O\_D(100) for \textbf{a} spin up, \textbf{b} spin down, and \textbf{c} heterojunction G/OH\_D(100) with degenerate spins. The Fermi level is set to 0. The Dirac point of graphene is marked by a blue dashed box. The computational setups and lattice structures can be found in the method section and supporting information.}
    \label{fig:band_dft}
\end{figure}

In the heterojunction between graphene and G/O\_D(100), we performed the spin-polarized calculations. The electron band structures with spin up and spin down are shown  in \cref{fig:band_dft}a and b, respectively. 
The Fermi energy level of the heterojunction is below the Dirac point of graphene, indicating that some electrons of graphene are transferred to O\_D(100). The charge difference plot in
the Supplementary Note 4 also confirms this finding, where 
we estimated 
the transferred electron density as 
$3 \times 10^{13}$ cm$^{-2}$. 
This value is larger than the estimated doping concentration in the Raman measurements, 
because our unit cell is small.
The L\"{o}wdin population analysis \cite{10.1063/1.1747632} also shows the atomic charge of graphene decreases from 3.956 to 3.949 electrons after patching the graphene
on O\_D(100). In summary, the observed hole doping is consistent with the G band and 2D band blue shift as found in the Raman measurements. The transferred electrons are paired with the surface unpaired electrons
on the diamond surface and thus decrease the spin noise. As a result, the $\mathrm{T_2}$ coherence time of the NV center increases. 

On the other hand, for the heterojunction G/OH\_D(100), we found that the Dirac point of graphene is very close to the Fermi energy level as shown in \cref{fig:band_dft}c,
indicating an insignificant charge transfer in the heterojunction G/OH\_D(100), which is consistent with the charge difference plot in the Supplementary Note 4. 
Therefore, our DFT calculations suggest that only choosing the right combination of patching materials and surface terminations can induce a favored charge transfer 
for electron pairing.


\subsection{Other surface treatment studies}\label{OH}

\begin{figure}[htbp]
    \centering
    \includegraphics[width=0.9\linewidth]{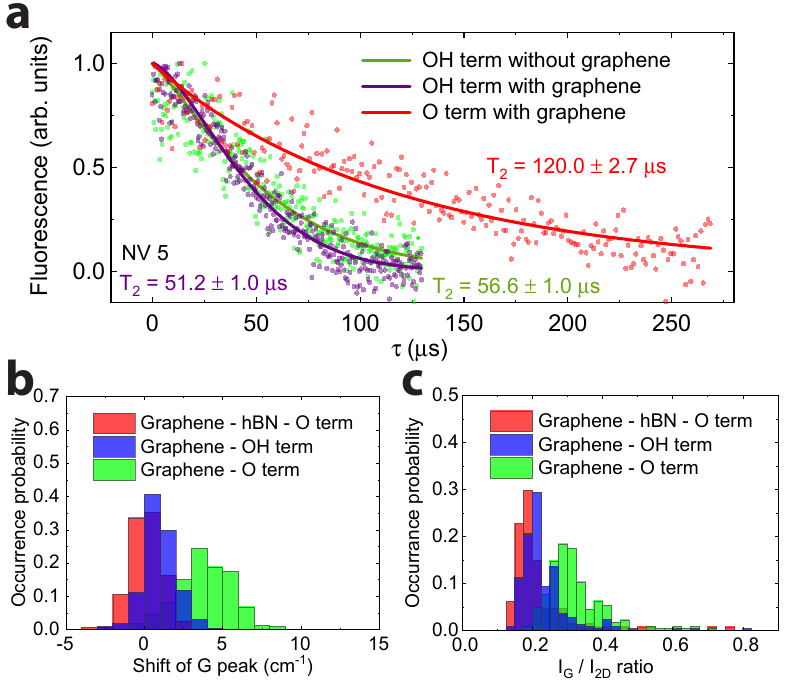}
    \caption{\textbf{Summary of $\mathrm{T_{2}}$ and Raman of OH-terminated diamond.} \textbf{a}, The Hahn-echo measurement of OH-terminated diamond. \textbf{b}, A statistic trend of the relative G band shift of graphene on different interfaces. The graphene on the h-BN flake at 1584 cm$^{-1}$ is set as the 0 shift reference. \textbf{c}, A statistic trend of the relative intensity $\mathrm{I_G/I_{2D}}$ ratio change on different interfaces. The histograms are collected by mapping the Raman spectra over a large area of heterojunction with around 200 data.}
    \label{fig:exp_OH}
\end{figure}

To experimentally validate the significance of surface terminations and patching materials, we conducted further investigations on other heterojunctions. Our DFT calculations in \cref{fig:band_dft} show that the OH-terminated diamond has a similar Fermi level to that of graphene, suggesting minimal electron transfer from graphene.
To confirm this theoretical prediction, we piranha-boil our diamond to create an OH-terminated diamond\cite{LI2021725} followed by the standard graphene transfer (see methods for details). In \cref{fig:exp_OH}a, the Hahn echo coherence time $\mathrm{T_2}$ of the NV center shows no significant improvement in the OH-terminated implanted diamond. The Raman spectroscopy results in \cref{fig:exp_OH}b and c exhibit a reduced blue shift of approximately 0.9 cm$^{-1}$ in the G band and less pronounced changes in the $\mathrm{I_G/I_{2D}}$ ratio, indicating reduced charge doping in graphene compared to O\_D(100). Since the OH-terminated surface has fewer unoccupied states that host unpaired electrons and more occupied OH functional groups, electron transfer from graphene is weakened, resulting in incomplete patching for G/OH\_D(100) which echoes with the DFT calculation results. Broadway et al. reported significant variations in diamond surface carrier density depending on cleaning and termination methods\cite{Broadway2018}. Therefore, graphene patching on an OH-terminated surface is incompetent in enhancing shallow spin coherence. Although the occupied OH functional groups exhibit less unpaired spin noise, they tend to absorb various molecules, introducing other noise sources to the NV and leading to shorter coherence time compared to graphene-patched O-terminated samples. 

Regarding patching materials, graphene, as a two-dimensional semi-metal, can shift its Fermi level to align with the contact material. Conversely, patching materials with a large band gap, such as h-BN ($\sim$5.6 eV), are ineffective as electron providers for pairing unpaired spins showing no enhancement of coherence time under h-BN patching. Therefore, integrating data from Raman spectroscopy, DFT, and NV measurements, we conclude that both surface termination and patching material are crucial for our interfacial engineering, which reduces unpaired electrons on the diamond surface and extends coherence time.

\subsection[Sensing demonstration - weakly coupled 13C]
      {Sensing demonstration - weakly coupled \ce{^{13}C}}\label{C13}

\begin{figure}[htbp]
   \centering
   \includegraphics[width=0.8\linewidth]{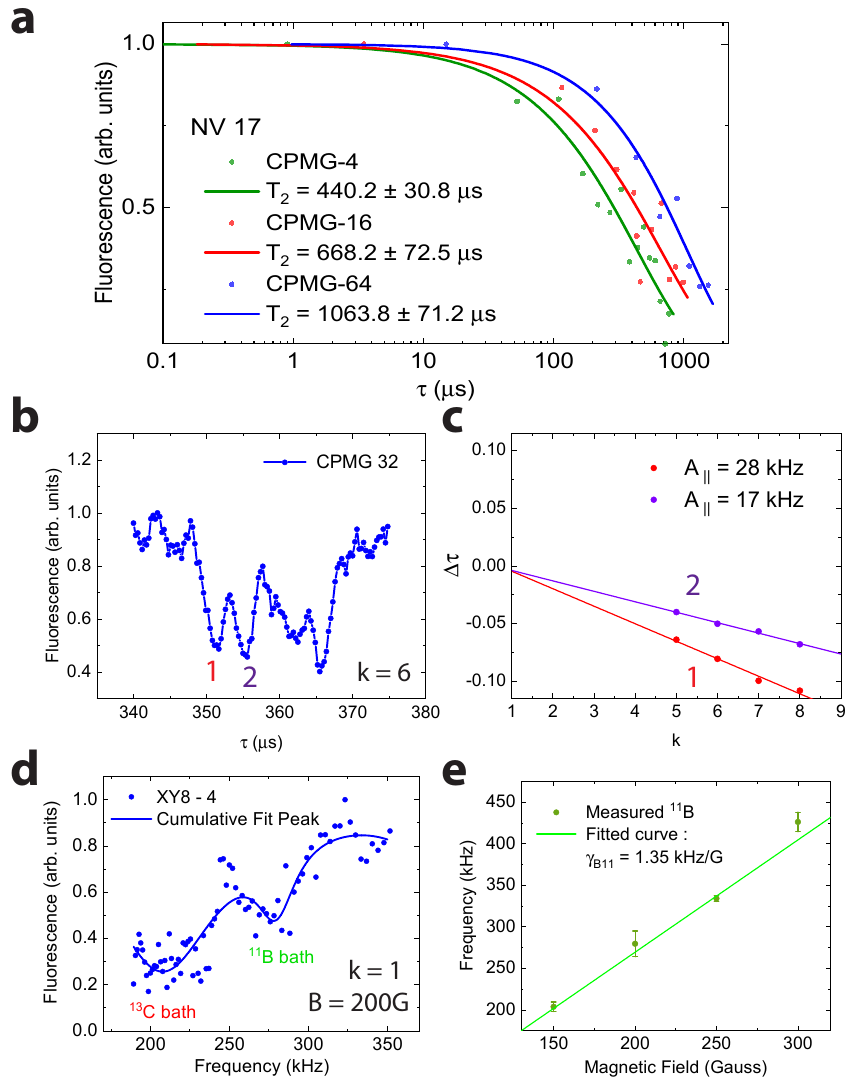}
   \caption{\textbf{Sensing of weakly coupled \ce{^{13}C} and \ce{^{11}B} spin bath by $T_2$ enhanced shallow NV center.} \textbf{a}, CPMG sequences with different numbers of $\pi$ pulses, N = 4, 16, 64 are applied to measure the coherence time of shallow NV centers in diamond. \textbf{b}, Coherence as a function of total evolution time for a decoupling sequence CPMG-32. The sharp peaks in the coherence signal correspond to coherent interactions with individual weakly coupled \ce{^{13}C} nuclear, where k=6 are the order of the resonance.
   \textbf{c}, The positions $\tau_k$ corresponding to the $k$-th order resonance peaks observed in \textbf{b}. $\Delta \tau=\tau_{k}/T_{L}-(2k-1)/4$ is the relative shift to the Larmor period $\mathrm{T_L}$ and the $\Delta \tau$ vs k confirms the resolved of weakly coupled \ce{^{13}C} spin. The coupling constants are resolved to be $A_\parallel$= 17 kHz/G and $A_\parallel$ = 28 kHz/G. \textbf{d}, XY8-4 sequence applied to measure the decoherence signal of the \ce{^{11}B} spin bath due to the h-BN capping layer. \textbf{e}, Tracking \ce{^{11}B} bath signal in XY8-4 with varying magnetic field magnitude B = 150, 200, 250, 300 G. Error bars represent the linewidth from the peak fitting.}
    \label{fig:C13}
\end{figure}

With greatly extended coherence time after our interfacial engineering, we study the sensitivity enhancement in shallow NV sensing. We first extend the coherence time further using CPMG dynamical decoupling techniques. In \cref{fig:C13}a, coherent time $\mathrm{T_2}$ of shallow NV center is further enhanced to above 1 ms. Our interface engineering method removes high-frequency electron noise on the surface, while the low-frequency noise from nuclear spins is filtered out by a dynamical decoupling sequence.  We then analyze the theoretical magnetic field sensitivity of the shallow NV centers to illustrate the improvement in sensing applications. The AC magnetic field sensitivity in echo measurements is given by the following equations \cite{Taylor2008,Herbschleb2019}:

\begin{align}
   \eta_{a.c.}\approx\frac{\pi\hbar}{2g\mu_B}\frac{\sqrt{\tau+t_{overhead}}}{C\tau}e^{(\tau/T_2)^p},
\end{align}

\begin{align}
   C=1/\sqrt{1+\frac{2(n_{0}+n_{1})}{(n_{0}-n_{1})^{2}}}.
\end{align}

Here, $\mathrm{C}$ represents the readout fidelity, which is calculated using the average photons ($n_{0}$ and $n_{1}$) of the spin states $m_{s}=0$ and $m_{s}=\pm1$. The typical value of $\mathrm{C}$ is 0.03 (100k counts/s and 30$\%$ contrast) when the NV centers are saturated \cite{doi:10.1126/science.aad8022}. Which $\mathrm{t_{overhead}}$ is the overhead time of the pulse sequence, $\tau$ is the free evolution interval, p is the fitted power of the exponent of the  $\mathrm{T_2}$.  Taking our recorded shallow NV center with the longest coherence time as an example  $\mathrm{T_2}$ enhanced from 212 $\mu s$ to 522 $\mu s$, fitted parameter p = 2.1. and $t_{overhead} = 3 \mu s$ , the AC magnetic field sensitivity at its best working $\tau$ improved from $50$ nTHz$^{-1/2}$ to $23$ nTHz$^{-1/2}$, resulting in a doubling of sensitivity, close to the best recorded AC magnetic field sensitivity of a single NV center in the bulk diamond, which has been reported to be 9 $\mathrm{nT/{Hz}^{1/2}}$\cite{Herbschleb2019}. However, to achieve such sensitivity required enriched $\mathrm{^{12}C}$ (99.998\%) CVD growth diamond, which required a series of complex fabrication process , while our device only required simple post-treatment. Upon the implementation of the CPMG-64 sequence, the sensitivity is further enhanced to $16$ nTHz$^{-1/2}$. By transferring graphene on nano-pillar\cite{doi:10.1021/acsnano.7b07689}, photon collection rates can be substantially enhanced ($>$ 300k counts/s \cite{PhysRevResearch.4.013098}), thereby increasing the $\mathrm{C}$ value and further improving sensitivity.

To gain further insights into how the reduction of surface electron noise benefits shallow NV centers in measuring weak magnetic signals, we demonstrate the measurement of weakly coupled $\mathrm{^{13}C}$ nuclear spins with our graphene patched O-terminated sample. Although some research has shown sensing of weakly coupled $\mathrm{^{13}C}$ nuclear spin with a deep NV center in bulk diamond with coupling strength ($\omega_h$) $\sim 2\pi\times$ 1 - 10 kHz \cite{PhysRevLett.109.137602, Abobeih2018}, the detection of weakly coupled $\mathrm{^{13}C}$ spins is limited in shallow NV centers due to constraints on their $\mathrm{T_2}$ times. With our interface engineering techniques, we demonstrated the observation of weakly coupled $\mathrm{^{13}C}$ with hyperfine strength ($\omega_h$) $\sim 2\pi\times$ 28 kHz and 17 kHz with $\sim 17$ nm shallow NV center at room temperature, which is comparable to the coupling strength sensed in bulk at cryogenic temperature with longer $\mathrm{T_1}$ of the NV electron spin \cite{Abobeih2018}. This is a dramatic improvement followed by enhanced electron spin coherence time. Here we choose N = 32 of the CPMG-N sequence to analyze the weakly coupled \ce{^{13}C} spin. We track the hyperfine peaks near the NV collapse for a few periods (k = 5, 6, 7, 8) and two \ce{^{13}C} hyperfine peaks can be identified \cite{PhysRevLett.109.137602}, which is $A_\parallel$ = 17 kHz and $A_\parallel$ = 28 kHz, see \cref{fig:C13}b and c. The coupling constant can be calculated by:

\begin{align}
   A_\parallel = \frac{(2k-1)\pi}{\tau_{k}}-2\omega_L,
\end{align}

Where $\tau_k$ is the position of the hyperfine peak and $\omega_L$ is the observed Larmor period. 

The detection of weakly coupled nuclear spins using interfacial engineered shallow NV centers demonstrates their potential for developing a reporter spin network capable of sensing, coherent coupling, and imaging individual proton spins on the diamond surface with nanometer 
resolution\cite{sushkov2014magnetic}.

\section[External sensing demonstration - 11B in h-BN] 
       {External sensing demonstration - \ce{^{11}B} in h-BN} \label{B11}

The ability to detect external nuclear spins is of paramount importance for nanoscale quantum sensing, as it unlocks the study of spin environments beyond the host crystal—such as molecules, 2D materials, and biological systems.
Measuring external spins, however, is significantly more challenging than detecting intrinsic spins like \ce{^{13}C}, as it requires a high-quality shallow NV center with long coherence time ($\mathrm{T_2}$) to measure the weak magnetic signal on the diamond surface. Lovchinsky et al.\cite{doi:10.1126/science.aal2538} achieved \ce{^{11}B} detection using isotopically purified (\ce{^{12}C}) 5-nm-deep NVs with 150 $\mu s$ $\mathrm{T_2}$ a feat demanding complex fabrication, including high-purity \ce{^{12}C} crystal growth and precision ion implantation, and NV pre-selection. While with graphene patching NVs, we are able to sense weak nuclear spins with 2 times greater sensing depth ( $\sim$ 10 nm) without any post-selection of the NV centers. To demonstrate our device's ability to perform external spin-sensing, we perform the measurement of \ce{^{11}B} nuclear spins in a few $\mu m $ h-BN sample transferred on top of our graphene-patched O-terminated sample, which we use as a capping layer to protect the hetero-junction.

The gyromagnetic ratio of \ce{^{11}B} is given by $\gamma_{B11}= 1.363 \mathrm{kHz/G}$. Therefore, the precession frequency of the \ce{^{11}B} nuclear spin is given by $f_{B11}=\gamma_{B11}B$. With our $\mathrm{T_2}$ enhanced shallow NV center, we drive the XY8-4 sequence to sense the magnetic signal given by the \ce{^{11}B} nuclear spin. We calculate the corresponding interrogation time of the \ce{^{11}B} bath as:

\begin{align}
   \tau = \frac{(2k-1)}{2\gamma_{B11}B} N_{\pi}
\end{align}

where k is the order of resonance and $N_{\pi}$ is the number of $\pi$ pulse in our sequence. In our study, we choose k = 1 and $N_{\pi} = 32$. Therefore, for example, in \cref{fig:C13}d with 200 G, the corresponding interrogation time for the decoherence to occur is 58.7 $\mu s$, corresponding to 272 kHz. We measured 279 kHz in this magnetic field, the discrepancy can be explained by the linewidth of the measurement, which is shown as the error in the measurement. We then vary the magnetic field (B = 150, 200, 250, and 300~G) and measure the corresponding signal. We can then fit the line of the measured frequency against magnetic field, and the fitted slope of the line is our measured $\gamma_{B11}$. The theoretical $\gamma_{B11}$ is given by $\mathrm{1.36~kHz/G}$ which is within the error of our measured $\gamma_{B11}$ which is $\mathrm{1.35 \pm 0.01~kHz/G}$ , see \cref{fig:C13}e.

Our results demonstrate that the enhanced sensitivity of the NV center enables external NMR measurements even with only an atomically thin graphene layer between the h-BN sample and the NV sensor.

\section[Discussion]
        {Discussion}\label{Discussion}

The interfacial engineering of shallow NV diamonds with graphene patching on O-terminated diamond represents a significant advance in the field of quantum sensing. By addressing the noise from surface unpaired spins, this method extends the coherence time of shallow NV centers to over 1 ms, nearing the $\mathrm{T_{1}}$ limit for NV centers. The use of Raman spectroscopy and DFT calculations to understand the role of diamond surface termination in triggering graphene charge transfer is particularly noteworthy. This approach not only achieves higher sensitivity but is also readily adapted to additional sensing applications. We have shown its ability to detect single weakly coupled \ce{^{13}C} nuclear spins and external \ce{^{11}B} from the capping h-BN.

\begin{figure}[htbp]
    \centering
\includegraphics[width=0.9\linewidth]{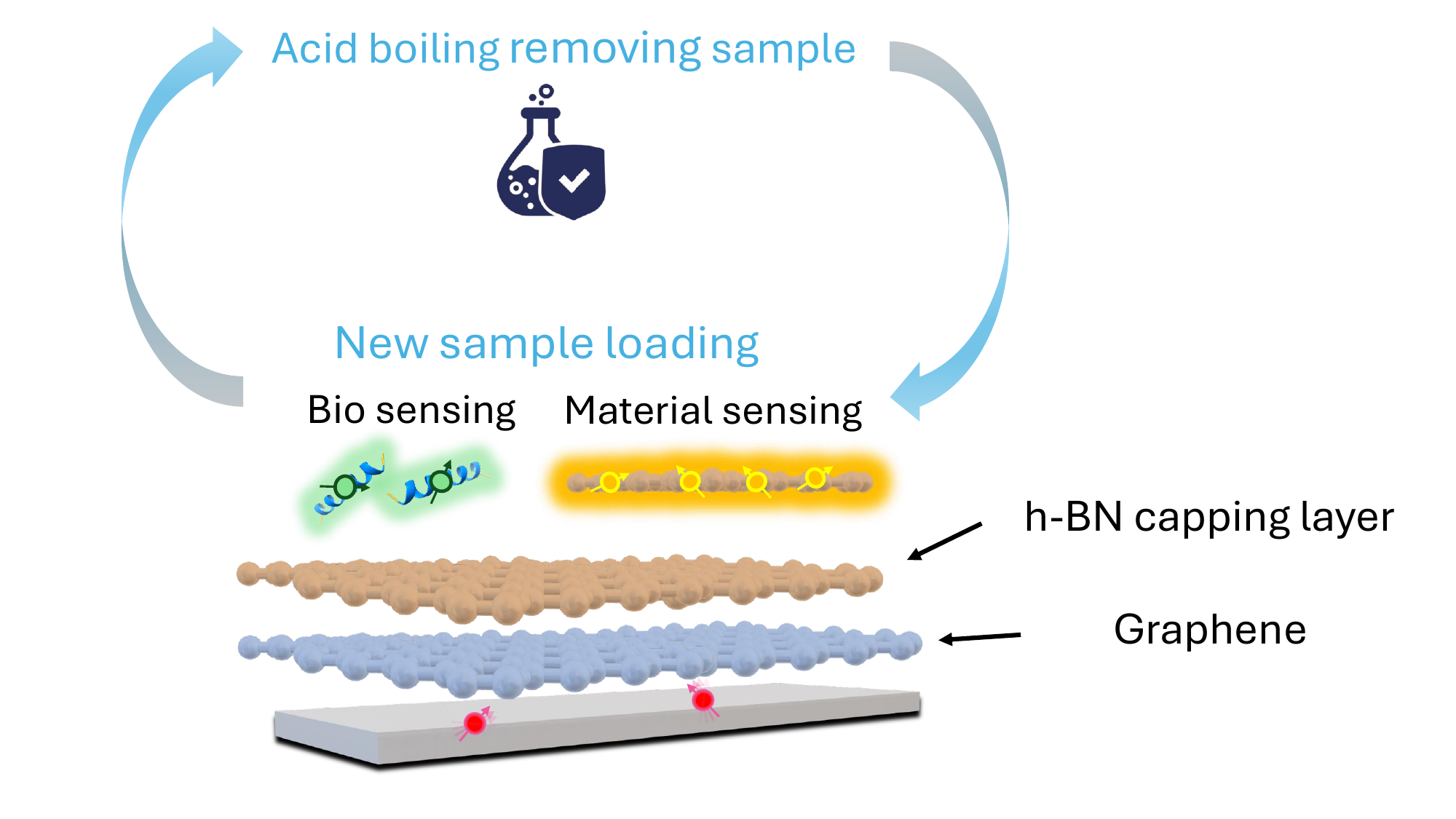}
    \caption{\textbf{Schematic diagram of h-BN-graphene diamond hetero-structure as a sensing device.} The h-BN-graphene-diamond hetero-structure can be repeatedly cleaned and reused for new experiments without removing the patching graphene.}
    \label{fig:exp_hBN}
\end{figure}

For practical deployment in quantum sensing, many NV optimization methods face practical trade-offs between performance and experimental flexibility. To facilitate practical usage, the device must prioritize straightforward sample loading and exchange. To addresses the critical need for both high sensitivity and sample accessibility, we designed a new hetero-structure as a device that integrates h-BN with our graphene-diamond hetero-structure (\cref{fig:exp_hBN}).  The acid resilient h-BN protects the graphene from strong oxidation during the common tri-acid cleaning, preserving coherence-enhancing properties. Therefore our h-BN-graphene-diamond device can be repeatedly cleaned and reloaded for new target sample without removing the patching graphene. This allows our device to be reused without compromising the enhancement effect. It serves a robust platform for near field measurement in most sensing scenarios including bio, high pressure, material study as shown in \cref{fig:exp_hBN}. This work paves the way for future research to facilitate the study of dilute protein assemblies\cite{shi2015single, doi:10.1126/science.aal2538}, DNA\cite{shi2018single} and paramagnetic species\cite{doi:10.1021/nl502988n} with single NVs and provides potential detection of exotic two-dimensional magnetic materials and their interactions at the atomic level. 

\section*{Methods}\label{Method}
\subsection*{Graphene transfer process}\label{ED:GTP}
The graphene is initially sandwiched between a sacrificial layer on top and a polymer layer at the bottom. The polymer layer is detached by floating the sacrificial-layer-graphene assembly in deionized water. Subsequently, the substrate (diamond or h-BN on diamond) is introduced into the deionized water, and the sacrificial-layer-graphene is positioned above it. The sacrificial-layer-graphene-substrate assembly is then air-dried for 30 minutes and annealed at 150°C for 1 hour. Following this, the device is stored under vacuum for a minimum of 24 hours to enhance the attachment of the sacrificial-layer-graphene. To remove the sacrificial layer, the device is immersed in 50\textsuperscript{o}C acetone for 1 hour, followed by immersion in isopropyl alcohol for an additional hour. The sample is then dried using nitrogen gas (\ce{N2}) and is ready for subsequent measurements.

\subsection*{NV spectroscopy}\label{ED:NV}
All $\mathrm{T_{2}}$ measurements are performed under a 286 G external magnetic field which is aligned to the 20 single NV in the implanted diamond. The same NVs are used to compare their performance with and without graphene and are confirmed by using confocal scanning microscopy. The rabi frequencies of the NV centers are fixed for a fair comparison (see Supplementary Note 1). The same 286 G external magnetic field and 7 NV (NV 1,2,3,5,7,16\&17) are used in DEER measurements. After locating the resonant frequency of the unpaired electrons, DEER rabi measurements are performed to further confirm the unpaired electron spin noise source \cite{PhysRevB.104.224412} and also to obtain the accurate time of flipping the unpaired electron spin ($t_{\pi}$), which allows us to perform DEER decay measurement (see Supplementary Note 1).  

\subsection*{Raman spectroscopy}
The Raman spectra were obtained by Renishaw inVia Qontor Spectrometer System. The data were obtained by a spectrometer with 600 lines/mm grating, 532 nm excitation, and a low power level (500 uW) to avoid any heating effect. The Raman spectra were then fitted with Gaussian functions. To obtain statistically meaningful data, Raman mapping was carried out in a region of 10 $\times$ 10 \textmu m$^{2}$ per each sample by scanning along x and y axes, thus providing around 200 independent probe spots.

\subsection*{Sample properties} \label{ED:Sample}
For single NV centers in implanted diamond, a 6 nm $p^{+}$ boron layer was first grown on the diamond and followed by a 9.8 keV \ce{^15N} ion implantation at a dose of approximately 10$^{9}$ N/cm\textsuperscript{2} and subsequent annealing at 950\textsuperscript{o}C for 2 hours. The boron-doped layer was then etched away by RIE O$_{2}$ plasma, in total 16 nm of diamond were etched. This technique can further enhance the $\mathrm{T_{2}}$ by removing multi-vacancy qubits in the diamond \cite{FávarodeOliveira2017}. NV is at a depth of approximately 5 - 20 nm below the diamond surface (see Supplementary Note 1). The diamond used in Raman measurements is the same implanted diamond sample. 

A home-built confocal microscopy is used for NV measurements. The objective is Olympus UMPlanFl, 100x 0.95NA. The laser power is 2 mW.

\subsection*{Surface termination} \label{ED:Surface termination}

The diamond sample is acid-boiled for oxygen termination(triacid boiling) and hydroxyl termination (piranha boiling)  before graphene transfer and optical measurements. Triacid cleaning uses a mixture of concentrated perchloric, nitric, and sulfuric acids to provide a range of oxygen functional groups on the surface of diamond sample\cite{PhysRevX.9.031052,REBULI19991620,LI2021725}. This is verified by XPS measurement. (see Supplementary Note 2) The detailed procedures of the triacid boiling is performed according to previous work\cite{BROWN201940}.  To create OH-terminated diamond, we use piranha solution (\ce{H_{2}O_{2}/H_{2}SO_{4}} mixture) to clean the diamond in 75\textsuperscript{o}C. With piranha cleaning, some dangling bonds of the diamond will be terminated by OH by forming C-OH \cite{Matsumae2019,LI2021725}.  It is confirmed with Attenuated Total Reflectance Fourier-transform infrared measurement (ATR-FTIR) as a stronger C-OH signal is observed when compared with the triacid cleaning method. (see Supplementary Note 2).

\subsection{DFT calculation}
DFT calculations were performed with the plane-wave pseudo-potential method implemented in the QUANTUM ESPRESSO package \cite{Giannozzi_2017}. We used the Perdew-Burke-Ernzerhof (PBE) \cite{PhysRevLett.77.3865} exchange-correlation (xc) functional, and the D3 van der Waals (vdW) \cite{10.1063/1.3382344, https://doi.org/10.1002/jcc.21759} corrections to model the long-range dispersion interactions. The kinetic energy cutoff for plane waves was 70 Ry. We used the G15 Optimized Norm-Conserving Vanderbilt (ONCV) \cite{hamann2013optimized,ONCV2} pseudopotentials. For G/O\_D(100), we did the spin-polarized calculation and used the Marzari-Vanderbilt (MV) smearing with a degauss value $10^{-4}$ Ry to broaden the occupation function near the Fermi level \cite{PhysRevLett.82.3296}.

A rectangular supercell of graphene containing 16 carbon atoms has the lattice dimensions of 2.46 $\text{\AA}$ and 17.05 $\text{\AA}$. 
The lattice dimension of the square unit cell of 
the diamond (100) surface slab is 2.52 $\text{\AA}$. 
To match the lattice of graphene, we used
a $(1\times7)$ diamond supercell. We compressed the diamond lattice by $2.38\%$ and 3.34\%, respectively, to match that of graphene. The Dirac point of rectangular graphene cell is located on the $\Gamma \rightarrow X$ path\cite{rectGraphene}. 
The diamond supercell contains 16 carbon layers.
We used a Monkhorst–Pack k-point mesh of $8\times1\times1$.
We optimized structures using convergence criteria for energy and force of $10^{-5}$ Ry and $10^{-4}$ Ry/Bohr, respectively. There is a 20 $\text{\AA}$ vacuum between neighboring slabs with periodic boundary conditions, and we further applied the dipole correction \cite{PhysRevB.59.12301}.

\backmatter

\subsection{Supplementary information}
Supplementary information is available.

\subsection{Data availability}
All data in this article are available from the corresponding authors upon request. Source data is available at https://doi.org/10.6084/m9.figshare.29092316.

\subsection{Acknowledgements}
S.Y. acknowledges financial support from Hong Kong Research Grants Council  (Projects RGC-AOE(AoE/P-701/20),  and GRF-16305422), NSFC (T2425002), and Guangdong Provincial Quantum Science Strategic Initiative (Grant No. GDZX2303005). D.P. acknowledges support from Hong Kong Research Grants Council (Projects GRF-16301723 and C6021-19EF).
During the manuscript writing, we noticed an independent work done by the Jiangfeng Du group in USTC, as referenced in National Science Review 12, nwaf076 (2025)\cite{10.1093/nsr/nwaf076}.

\subsection{Author contribution}
S.Y. and D.P. designed and supervised the project. J.W., H.Y.C. and Y.S. constructed the experimental 
setup. H.Y.C., W.K.L. and J.W. performed the experiments and data acquisition. Y.Z. performed the DFT calculations and theoretical interpretation. W.K.L., Y.C., H.Y.C. and J.W. prepared the experiment samples. W.K.L., Y.Z., H.Y.C., J.W., K.O.H., M.Y.L., X.D., S.Y. and D.P performed the data analysis and interpretation. K.O.H., H.Y.C., J.W., Y.Z. and W.K.L. wrote the article, with input from all other authors. All the authors commented on the final article.

\subsection{Competing interests}
The authors declare no competing interests.

\end{document}